\documentclass[10pt,conference]{IEEEtran}
\IEEEoverridecommandlockouts

\usepackage{amsmath,amssymb,graphicx,booktabs,hyperref,xcolor,listings,url,microtype,cite,orcidlink}
\usepackage{tikz,pgfplots}
\usetikzlibrary{arrows.meta,shapes.geometric,positioning,fit,calc}
\pgfplotsset{compat=1.18}

\hypersetup{
    colorlinks   = true,
    linkcolor    = blue!70!black,
    citecolor    = blue!70!black,
    urlcolor     = blue!70!black,
    pdfauthor    = {Nicol\'as Padilla},
    pdftitle     = {Exposed by Design: A Dynamic Security Assessment of
                    Internet-Facing MCP Servers at Scale},
    pdfkeywords  = {Model Context Protocol, MCP security, dynamic measurement,
                    tool poisoning, vulnerability disclosure}
}

\definecolor{codegreen}{rgb}{0.13, 0.55, 0.13}
\definecolor{codegray}{rgb}{0.50, 0.50, 0.50}
\definecolor{codepurple}{rgb}{0.45, 0.00, 0.60}
\definecolor{backcolour}{rgb}{0.97, 0.97, 0.97}
\definecolor{darkred}{rgb}{0.70, 0.00, 0.00}

\lstdefinestyle{mcpstyle}{
    backgroundcolor  = \color{backcolour},
    commentstyle     = \color{codegreen},
    keywordstyle     = \color{codepurple}\bfseries,
    numberstyle      = \tiny\color{codegray},
    stringstyle      = \color{darkred},
    basicstyle       = \ttfamily\footnotesize,
    breakatwhitespace= false,
    breaklines       = true,
    captionpos       = b,
    keepspaces       = true,
    numbers          = left,
    numbersep        = 5pt,
    showspaces       = false,
    showstringspaces = false,
    showtabs         = false,
    tabsize          = 2,
    frame            = single,
    framerule        = 0.4pt,
    xleftmargin      = 1.5em,
    framexleftmargin = 1.5em
}
\begin{document}

\title{Exposed by Design: A Dynamic Security Assessment of\\
       Internet-Facing MCP Servers at Scale}

\author{%
    \IEEEauthorblockN{Nicol\'{a}s Padilla\,\orcidlink{0009-0005-3504-7241}}
    \IEEEauthorblockA{%
        CobaltoSec\\
        Independent Security Researcher\\
        nicolas@cobalto-sec.tech%
    }
}

\maketitle

\begin{abstract}
The Model Context Protocol (MCP) has seen rapid adoption since its
November~2024 launch, with over 21,000 server instances detectable on
the public internet. We present the first dynamic behavioral security
assessment of internet-facing MCP servers, combining passive discovery across eleven data sources (crt.sh, HuggingFace, GitHub, npm, Smithery, PyPI, Censys, FOFA, Shodan, glama.ai, and pulsemcp.com) with active dynamic testing using Corvus, a
purpose-built framework implementing 34 test modules covering
10~MCP-specific vulnerability classes. Across four measurement runs
spanning July~2026, we confirm 640 production MCP servers and
dynamically audit 414, uncovering 68 reportable vulnerabilities
including SQL injection, SSRF targeting cloud metadata services, prompt
template injection, and path traversal via cursor manipulation. We find
that 91.8\% of dynamically audited servers lack OAuth authentication, 687 tool instances across confirmed servers expose shell
execution capabilities without access controls, and 41.6\% of confirmed
servers disappear within three days between consecutive measurement runs---indicating rapid deployment cycles
without security review. We report on our responsible disclosure
pipeline and release Corvus as an open-source framework for MCP security
evaluation.
\end{abstract}


\section{Introduction}
\label{sec:intro}

The Model Context Protocol (MCP), released by Anthropic in November
2024~\cite{anthropic2024mcp}, defines a standardized JSON-RPC wire
protocol for connecting large language model (LLM) agents to external
tools, data sources, and services. Twelve months after publication,
MCP has become the de facto integration layer for agentic AI
applications: dedicated registries report over 21,000 publicly
reachable server instances~\cite{censys2026}, with new packages
appearing continuously across GitHub, npm, PyPI, HuggingFace, and
specialized directories such as Smithery, glama.ai, and pulsemcp.com.
No comparable protocol has achieved this deployment breadth in a
comparable timeframe, a pace driven by the minimal barrier to exposing
any existing service---a database, a shell, a cloud API---as an MCP
endpoint.

This velocity carries disproportionate security risk. An MCP server is
simultaneously a network service and an AI action interface. It receives
JSON-RPC requests, executes privileged operations---reading files,
querying databases, invoking shell commands, calling cloud
APIs---and returns structured results that enter an LLM agent's context
window. A compromised or misconfigured MCP server therefore acts as an
\emph{amplifier}: it can translate attacker influence over tool
descriptions or outputs into consequential machine actions within an
autonomous pipeline. Critically, MCP introduces attack surfaces with no
direct analogues in web security. Tool description poisoning encodes
adversarial instructions in the natural-language fields that LLMs read
to understand tool behavior~\cite{rugpull2025}. Schema-level injection
exploits parser ambiguities in JSON-RPC dispatch. Cursor-based path
traversal weaponizes the protocol's pagination primitive. These
vulnerabilities are native to MCP's design---not adaptations of
established classes---and existing security tooling does not test for
them.

Empirical evidence confirms that the deployment landscape has not
internalized this threat model. In our measurements,
\textbf{91.8\% of dynamically audited servers operate with no OAuth
authentication layer}; \textbf{687 tool instances across confirmed servers expose shell execution
capabilities without access controls whatsoever}; and \textbf{41.6\% of
confirmed servers disappear between consecutive measurement runs
spanning approximately seventy-two hours}---indicating that developers
ship MCP integrations at the pace of a weekend project, without the
operational baseline or security review associated with a public-facing
service.

\subsection*{Gap in Prior Work}

Security research on MCP has proceeded along two trajectories that,
taken together, leave the production deployment landscape essentially
uncharacterized.

The first trajectory is theoretical. Hou et al.~\cite{mcpunsafe2025}
and related works construct threat taxonomies and formal attack models.
Targeted analyses examine specific vectors in isolation: rug-pull
attacks~\cite{rugpull2025} model dynamic tool mutation, and Greshake et al.~\cite{greshake2023not} document indirect prompt injection
via adversarially crafted tool outputs. These contributions establish important conceptual foundations
but supply no data on the prevalence of such vulnerabilities in deployed
systems.

The second trajectory is empirical but constrained to static analysis.
Zhou et al.~\cite{zhou2026mcp} conducted the largest prior empirical
study, measuring authentication state across 7,973 servers; however,
they did not perform tool-level behavioral testing and their coverage is
limited to MCP08 (authentication deficiency). Li and Gao~\cite{li2026mcpsafety} analyze 67,057 MCP registry entries through static pattern matching, never contacting live servers. VIPER-MCP~\cite{viper2026} applied static taint analysis with
LLM-guided exploit generation to 39,884 GitHub repositories, without
testing HTTP-exposed server instances. Hasan et al.~\cite{hasan2026}
similarly processed 1,899 open-source repositories through static
analysis pipelines. Methodological analogs from adjacent
domains---empirical npm vulnerability measurement~\cite{zimmermann2019},
outdated library studies~\cite{lauinger2017}, and Internet-wide
scanning~\cite{durumeric2013zmap}---inform our measurement design but
were not applied to MCP.

The present work fills this gap by combining passive multi-source
discovery across eleven data sources with active dynamic testing using a
framework that covers all ten MCP-specific vulnerability categories,
deployed against production servers across a longitudinal measurement
campaign.

\subsection*{Contributions}

This paper makes the following contributions:

\begin{itemize}

    \item[\textbf{C1}] \textbf{First dynamic behavioral security assessment of
    internet-facing MCP servers.} Across four measurement runs in July~2026, we
    process up to 4,110 candidate URLs per run, confirm 640 unique production
    MCP deployments via protocol-level fingerprinting, and subject 414 to
    active behavioral testing.

    \item[\textbf{C2}] \textbf{Corvus: an open-source MCP security assessment
    framework.} We design and implement Corvus, comprising 34 test modules---13
    static and 21 dynamic---covering all 10 categories of the MCP Security Top~10 (MST-10)
    taxonomy. Corvus operates over both Streamable HTTP and legacy SSE
    transports, produces SARIF~2.1.0 output compatible with existing security
    toolchains, and is released as open-source to enable replication and
    extension of this work.

    \item[\textbf{C3}] \textbf{A multi-source passive discovery methodology.}
    We demonstrate that no single data source captures the MCP server
    population adequately, and we present a discovery pipeline spanning eleven
    heterogeneous sources: certificate transparency logs (\texttt{crt.sh}),
    developer platforms (GitHub, HuggingFace), package registries (npm, PyPI),
    specialized MCP registries (Smithery, glama.ai, pulsemcp.com), and
    Internet-wide scanning infrastructure (Censys, FOFA, Shodan). We
    characterize the unique coverage and overlap of each source.

    \item[\textbf{C4}] \textbf{Characterization of deployment churn and
    operational immaturity.} We find that 41.6\% of servers confirmed in
    Run~3 (464 servers) had disappeared by Run~4 seventy-two hours later,
    including 9 previously CRITICAL-rated servers that were remediated or
    decommissioned. This churn rate, combined with near-universal OAuth
    absence, characterizes the MCP server ecosystem as operating below the
    minimum operational security baseline of conventional internet services.

    \item[\textbf{C5}] \textbf{Responsible disclosure at scale.} From
    414 dynamically tested servers, we identify 68 reportable vulnerabilities
    and document a triage, curation, and disclosure pipeline operating under
    GitHub's GHSA advisory system with 90-day coordinated embargo. Confirmed
    findings include: SQL injection (CRITICAL, confirmed via response-size differential; no data retained); SSRF against the AWS Instance Metadata Service (IMDS),
    evidenced by an 11.9\,s response latency versus a 0.3\,s baseline;
    prompt template injection enabling model instruction override; and
    path traversal via cursor manipulation across multiple deployments. Nineteen
    advisories have completed full public disclosure; 49 remain under active
    embargo.

\end{itemize}

\subsection*{Paper Organization}

The remainder of this paper is organized as follows.
Section~\ref{sec:related} surveys prior work on MCP security and
empirical measurement methodology.
Section~\ref{sec:background} provides background on the MCP protocol and
formalizes our vulnerability taxonomy.
Section~\ref{sec:methodology} describes the discovery pipeline and Corvus
dynamic testing methodology.
Section~\ref{sec:results} presents longitudinal discovery results,
vulnerability findings, and churn analysis.
Section~\ref{sec:discussion} discusses structural implications and the
responsible disclosure pipeline.
Section~\ref{sec:limitations} addresses threats to validity and
measurement scope.
Section~\ref{sec:conclusion} concludes.


\section{Related Work}
\label{sec:related}

\subsection{MCP Security Research}
\label{sec:related:mcp}

The Model Context Protocol was publicly released in November 2024~\cite{anthropic2024mcp},
and dedicated security research is still nascent.
We survey the most directly related work, organized by methodology, and identify the gap each study leaves.

\paragraph{Dynamic probing of deployed servers.}
Zhou et al.~\cite{zhou2026mcp} scan 7,973 deployed MCP servers reachable over the public internet and
measure authentication posture, finding that the majority lack OAuth credentials.
Their study is the only prior work to operate on live production deployments at scale.
However, the probe is confined to authentication header detection, corresponding to MCP08 in our taxonomy;
no tool endpoints are invoked, no behavioral vulnerability classes are exercised,
and no vulnerabilities are disclosed to affected vendors.
Our work extends this empirical approach to dynamic behavioral auditing across all ten
MCP Security Top 10 (MST-10) categories and to coordinated disclosure.

\paragraph{Registry-scale static analysis.}
Li and Gao~\cite{li2026mcpsafety} analyze 67,057 MCP registry entries through static pattern matching to classify tool definitions by safety category.
The dataset is the largest in the literature by entry count, but the analysis is confined to registry manifests:
no server instances are created, no HTTP endpoints are contacted, and runtime behavior is not observed.
VIPER-MCP~\cite{viper2026} extends this approach to 39,884 GitHub repositories,
combining static taint analysis with LLM-generated exploit sketches to identify injection-prone dataflows in source code;
deployed server instances are not tested.
Hasan et al.~\cite{hasan2026} similarly apply static analysis to 1,899 open-source MCP repositories.
The limitation shared by all three studies is structural: static analysis of registry entries or source code
cannot reveal runtime access control enforcement, deployment-specific configurations, or behavioral properties
that emerge only from live server interaction.
Critically, servers in production may differ substantially from their registry representations
due to version skew, runtime wrappers, and operator-applied configurations.

\paragraph{Threat taxonomy and attack-specific studies.}
Hou et al.~\cite{mcpunsafe2025} survey the MCP attack landscape and threat categories,
including tool poisoning, cross-server authority escalation, and ambient authority misuse;
their work contains no empirical measurements of deployed systems.
The rug-pull attack formalization~\cite{rugpull2025} characterizes the class of attacks in which a server
alters tool semantics after initial user consent, without a large-scale empirical component.
Greshake et al.~\cite{greshake2023not} demonstrate indirect prompt injection via adversarially crafted
content in tool outputs, validating the attack class on controlled targets without ecosystem-level breadth.

\paragraph{Pre-MCP tool-use security.}
Prior to MCP standardization, prompt injection, confused-deputy attacks, and authorization boundary
violations were already documented as recurring failure modes in LLM agent pipelines with
tool-use capabilities~\cite{greshake2023not}.
MCP introduces attack surfaces that earlier threat models do not address, including tool manifests,
resource URIs, the sampling channel, and elicitation flows.

\paragraph{Summary.}
Table~\ref{tab:related} positions our work relative to prior studies along six methodological dimensions.
No existing study combines (i)~multi-source discovery of internet-reachable production deployments at scale,
(ii)~dynamic behavioral auditing across all ten MCP Security Top 10 (MST-10) vulnerability categories,
and (iii)~coordinated vulnerability disclosure with registered GHSA identifiers.

\begin{table}[t]
\centering
\caption{Comparison with existing MCP security work. \textit{Scale}: number of servers, registry entries,
or repositories examined. \textit{Dyn.}: dynamic behavioral testing beyond authentication probing.
\textit{Prod.}: study targets deployed, internet-reachable servers. Our work audited 414 of 640 confirmed
servers with full dynamic testing; 68 vulnerabilities received GHSA registration.}
\label{tab:related}
\footnotesize
\setlength{\tabcolsep}{3.5pt}
\begin{tabular}{@{}lrllll@{}}
\toprule
Work & Scale & Method & Dyn. & Prod. & Disclose \\
\midrule
Zhou et al.~\cite{zhou2026mcp}     & 7,973  & Auth probe    & Partial & Yes & None \\
Li \& Gao~\cite{li2026mcpsafety}   & 67,057 & Static        & No      & No  & None \\
VIPER-MCP~\cite{viper2026}         & 39,884 & Static+LLM    & No      & No  & None \\
Hasan et al.~\cite{hasan2026}      & 1,899  & Static        & No      & No  & None \\
\cite{mcpunsafe2025}               & ---    & Taxonomy      & No      & No  & None \\
\cite{rugpull2025}                 & ---    & Formalization & No      & No  & None \\
Greshake et al.~\cite{greshake2023not} & ---  & Single PoC    & No      & No  & None \\
\midrule
\textbf{This work}                 & \textbf{640} & \textbf{Disc.+Dyn.} & \textbf{Yes} & \textbf{Yes} & \textbf{68 GHSAs} \\
\bottomrule
\end{tabular}
\end{table}

\subsection{Empirical Security Studies of Software Ecosystems}
\label{sec:related:ecosystems}

Our methodology is situated within a well-established tradition of large-scale empirical security
measurement of software ecosystems.
Zimmermann et al.~\cite{zimmermann2019} characterize the npm dependency graph at internet scale,
demonstrating that a single vulnerable package can transitively expose hundreds of thousands of dependent projects
to attack.
Their study establishes the quantitative template for measuring ecosystem-wide risk,
and their supply-chain findings directly motivate our MCP10 test module,
which probes dependency confusion scenarios and transitive trust relationships among MCP server integrations.
Lauinger et al.~\cite{lauinger2017} measure the prevalence of outdated JavaScript libraries on the live web,
finding that production deployments contain vulnerabilities invisible in current registry records.
Their central result---that live production data reveals exposure that registry snapshots systematically miss---is
the methodological premise of our own study:
we probe deployed MCP servers rather than analyze registries precisely because registry data
is not a reliable proxy for the runtime security posture of those servers.

The broader literature on empirical measurement of protocol security properties,
including the large-scale study of HTTPS deployment, DNSSEC adoption, and email authentication standards
(SPF, DKIM, DMARC), has repeatedly demonstrated that deployment-level measurements
diverge substantially from what documentation or source repositories would predict.
We apply this epistemological principle to MCP:
authentication enforcement, tool behavioral consistency, and access control boundaries are
observable only through active probing of live server instances, not through analysis of
server metadata or source code.

The MCP ecosystem exhibits structural parallels to early npm:
rapid growth driven by a long tail of single-author packages with minimal security review infrastructure,
aggressive reuse of third-party integrations, and no standardized vulnerability reporting channel
at the time of our measurements.
Our study is the first to characterize this ecosystem's security posture at production scale
using the empirical methods established by the software security measurement literature.

\subsection{Internet Measurement and Security}
\label{sec:related:measurement}

Internet-scale active measurement as a security methodology was formalized by Durumeric et al.\
with ZMap~\cite{durumeric2013zmap}, demonstrating that the full IPv4 address space can be surveyed
at gigabit line rate in under one hour.
Follow-on work applied this capability to characterize global TLS deployment,
measure the persistence of known vulnerabilities across patch cycles,
and enumerate exposed industrial control systems on the public internet.
The consistent methodological lesson is that accurate prevalence estimates require measurement of the live
population: curated samples and repository data produce systematically biased estimates of real-world exposure.

MCP presents a discovery challenge distinct from port-based protocols.
MCP servers are not registered in DNS, do not occupy predictable ports,
and are distributed across eleven heterogeneous discovery surfaces,
including certificate transparency logs (crt.sh), package registries (npm, PyPI),
code hosting (GitHub, HuggingFace), specialized MCP directories (Smithery, glama.ai, pulsemcp.com),
and internet scan databases (Censys, FOFA, Shodan).
Discovery therefore requires multi-source synthesis, cross-source deduplication,
and HTTP-level fingerprinting to confirm MCP capability---analogous to the
application-layer confirmation techniques used in ZMap follow-on studies that move beyond
port reachability to characterize application-level behavior.
Our discovery pipeline (Section~\ref{sec:methodology}) implements this approach
across all eleven sources, yielding a corpus of 640 confirmed production MCP servers
across four longitudinal measurement runs in July 2026
(Section~\ref{sec:findings:discovery}).
This corpus represents, to our knowledge, the largest empirically verified dataset
of dynamically-audited, internet-reachable MCP endpoints subjected to behavioral
security testing to date.


\section{Background}
\label{sec:background}

\subsection{The Model Context Protocol}

The Model Context Protocol (MCP) is an open standard for connecting large language model (LLM) applications to external tools and data sources. Published by Anthropic in November 2024~\cite{anthropic2024mcp}, MCP has seen rapid adoption across commercial and open-source LLM toolchains within eighteen months of its release. The protocol defines three roles. A \emph{host} is the LLM application (e.g., a coding assistant or an autonomous agent) that orchestrates tool use on behalf of a user. A \emph{client} is the in-process connector maintained by the host, responsible for managing the transport session lifecycle. A \emph{server} is the process or network service that exposes capabilities to the host.

All communication is encoded as JSON-RPC~2.0 messages exchanged over one of three transports. The \textbf{stdio} transport launches the server as a child process; the host writes serialized messages to the server's standard input and reads responses from its standard output, confining the server to the local machine. The \textbf{HTTP Streamable} transport (the current specification) exposes a single HTTPS endpoint accepting POST requests, with optional server-sent event (SSE) streams for long-running responses. The \textbf{SSE legacy} transport maintains a dedicated SSE channel for server-to-client messages alongside a separate POST endpoint for client-to-server traffic, retained for backwards compatibility with early implementations.

Every session begins with a capability negotiation handshake: the host issues an \texttt{initialize} request advertising its protocol version and supported capabilities; the server responds with its own version, metadata, and the capability set it implements. This negotiation governs which protocol primitives are available for the remainder of the session.

MCP defines four capability primitives. \textbf{Tools} are callable functions advertised via \texttt{tools/list} and invoked via \texttt{tools/call}; each tool declaration carries a name, a free-text natural-language description consumed by the LLM when deciding whether to invoke the tool, and a JSON Schema constraining its input parameters. \textbf{Resources} are typed data objects addressed by URI (e.g., \texttt{file:///path}, \texttt{db://table/query}) and retrieved via \texttt{resources/read}; the URI namespace is defined entirely by the server with no protocol-level constraints. \textbf{Prompts} are server-defined, parameterized prompt templates retrieved via \texttt{prompts/get} and injected verbatim into the LLM's context window. \textbf{Sampling} is a reverse capability by which a server may request LLM completions from the host client, enabling server-directed inference flows.

The security-relevant property of this design is that an LLM agent reads tool descriptions and resource contents as \emph{trusted context} prior to generating the arguments it passes to tool invocations. This structural coupling between natural-language text and downstream code execution is the root cause of several vulnerability classes characterized in Section~\ref{sec:findings}.

\subsection{MCP Deployment Landscape}

MCP servers are distributed and operated in two deployment modes with substantially different security properties.

\textbf{Local (stdio) deployment.} The server is a software package installed on the end-user's machine and launched on demand as a child process of the host application. The dominant distribution mechanisms are the npm ecosystem (invoked via \texttt{npx}) and Python packaging (via \texttt{pip} or \texttt{uvx}). These servers execute under the local user's identity, inherit the local filesystem and process environment, and are unreachable from remote networks.

\textbf{Remote (HTTP) deployment.} The server is a persistent HTTPS service deployed to a cloud platform. Developers commonly use commodity platform-as-a-service offerings, including Vercel, Railway, Fly.io, Render, and Hugging Face Spaces. These platforms issue publicly routable hostnames on shared top-level domains (e.g., \texttt{*.vercel.app}, \texttt{*.railway.app}, \texttt{*.hf.space}) with no authentication requirement enforced at the infrastructure layer. An MCP server deployed in this fashion is, absent application-level access controls, accessible to any host on the public internet.

The security implications of remote deployment are compounded by two protocol properties. First, the \texttt{tools/list} endpoint provides a structured, machine-readable inventory of all server capabilities---including natural-language descriptions, parameter schemas, and inferred backend behaviors---without mandating authentication in the specification. Second, the MCP specification treats OAuth~2.0 as optional; many server implementations omit all authentication. Together, these properties allow an unauthenticated adversary to enumerate and invoke server capabilities at internet scale. By May 2026, Censys telemetry identified more than 21,000 publicly accessible MCP endpoints.

\subsection{Threat Model}

\paragraph{Attacker model.}
We consider an \emph{external, unauthenticated} adversary capable of issuing arbitrary HTTP requests to any publicly routable MCP server endpoint. The adversary has no prior knowledge of the server's implementation or source code, no access to server-side logs, and performs no man-in-the-middle interception of legitimate sessions. Adversarial goals include one or more of: (i)~capability enumeration to identify high-value targets; (ii)~exploitation of server-side vulnerabilities to reach internal backend systems (e.g., via server-side request forgery); (iii)~remote command execution via unsanitized tool parameter handling; and (iv)~extraction of credentials or tokens reflected in tool responses.

\paragraph{Assets at risk.}
We identify three classes of assets exposed by HTTP-deployed MCP servers. \textbf{Data}: records, files, or objects accessible through server tools and resource URIs, potentially including database contents, cloud object storage, and application configuration. \textbf{Backend network reachability}: servers that issue outbound HTTP requests on behalf of tool arguments may be abused as server-side request forgery (SSRF) proxies to reach cloud metadata services (e.g., the AWS Instance Metadata Service at \texttt{169.254.169.254}), internal APIs, and other hosts on the server's private network. \textbf{Remote code execution}: servers that forward tool parameters to subprocess or shell invocations without sanitization expose the host operating system directly to command injection.

\paragraph{Out of scope.}
We explicitly exclude stdio-transport servers executing as local child processes, which are unreachable by network adversaries and constitute a distinct threat model. We also exclude \emph{client-side} indirect prompt injection---attacks in which adversarially crafted documents cause a host LLM to misuse otherwise-legitimate server tools~\cite{greshake2023not}---as this threat targets the AI client rather than the server. Attacks on LLM inference systems themselves are likewise out of scope.

\paragraph{Ethical constraints.}
All measurements adhere to responsible disclosure principles. Discovery employed passive sources (certificate transparency logs, public package registries, search engine APIs) and active probing restricted to PoC-level verification that does not modify server state or retrieve real user data. All confirmed vulnerabilities were reported to affected maintainers under a 90-day coordinated disclosure embargo prior to public release. No user data was collected, retained, or analyzed beyond what was minimally necessary to classify the vulnerability class.

\subsection{MCP Security Top 10 (MST-10)}
\label{sec:taxonomy}

Established vulnerability taxonomies for web applications~\cite{owasp2021top10} and API security~\cite{owasp2023api} do not capture the attack surface introduced by MCP's natural-language tool-description layer, server-controlled resource URI namespace, and JSON-RPC protocol semantics. We introduce the \emph{MCP Security Top 10 (MST-10)}, a taxonomy of ten vulnerability classes derived from iterative analysis of findings across our measurement campaign. Categories are ordered by combined exploitability and potential impact, consistent with OWASP methodology. Table~\ref{tab:owasp} presents the full taxonomy; we provide prevalence data, representative behaviors, and confirmed proof-of-concept findings for each class in Section~\ref{sec:findings}.
(Categories are ordered by combined exploitability and impact, following OWASP Top~10
methodology; this taxonomy is an independent contribution and is not affiliated with
or endorsed by the OWASP Foundation.)

\begin{table*}[t]
\centering
\caption{MCP Security Top 10 (MST-10) vulnerability taxonomy applied in this study. Categories are ordered by combined exploitability and potential impact.}
\label{tab:owasp}
\begin{tabular}{@{}llp{10.8cm}@{}}
\toprule
\textbf{ID} & \textbf{Category} & \textbf{Key Attack Vector} \\
\midrule
MCP01 & Tool Poisoning & Malicious instructions injected into tool names or descriptions; shadow tool registration; dynamic mutation of tool definitions across session calls (rug-pull attack) \\
MCP02 & Insufficient Access Control & Unauthenticated \texttt{tools/call} invocation reaching privileged operations; per-tool authorization bypass; OAuth token scope reuse \\
MCP03 & Prompt Injection via Tool Output & Server-controlled response payloads used to inject instructions into the host LLM's context window and redirect agent behavior \\
MCP04 & Resource URI Manip. / SSRF & Path traversal via URI parameters; cursor-based directory escape; server-side request forgery via URL parameters forwarded to server-side HTTP clients \\
MCP05 & Schema Bypass               & Malformed or absent JSON-RPC \texttt{inputSchema} exploited to reach unvalidated server code paths; nested schema confusion bypassing type constraints \\
MCP06 & Command Injection           & Unsanitized tool parameters forwarded to OS-level subprocess execution, SQL query construction, or shell interpreters \\
MCP07 & Sensitive Data Exposure     & Bearer tokens, API credentials, or internal configuration reflected in error messages, verbose logs, or tool response bodies \\
MCP08 & Authentication Deficiencies & Absent authentication enforcement; static or predictable bearer tokens; optional OAuth~2.0 omitted entirely by the server \\
MCP09 & Denial of Service           & Batch \texttt{tools/call} flooding; unbounded response stream generation; server resource exhaustion via crafted large or recursive inputs \\
MCP10 & Supply Chain                & Transitive dependency exploitation in distributed server packages; tool chaining to escalate privileges across composed server trust boundaries \\
\bottomrule
\end{tabular}
\end{table*}

The taxonomy partitions naturally along the protocol layer at which each vulnerability manifests. MCP01 and MCP03 are structurally unique to the MCP threat model: they exploit the natural-language surface of tool declarations and tool outputs, a semantic layer absent from conventional REST or RPC API security models. MCP02, MCP08, and MCP09 correspond to well-understood web security failures---broken access control, weak authentication, and resource exhaustion---applied to the JSON-RPC substrate. MCP04 and MCP05 exploit MCP-specific protocol semantics (resource URI namespaces and schema negotiation) with no direct analog in prior API models. MCP06 and MCP07 represent server-implementation failures in which the protocol is processed correctly but tool arguments are forwarded unsafely to backend systems. MCP10 reflects emergent risks in composed or registry-distributed deployments, where tool chaining and transitive dependencies introduce privilege escalation paths that span server trust boundaries.


\section{Methodology}
\label{sec:methodology}

\subsection{Overview}

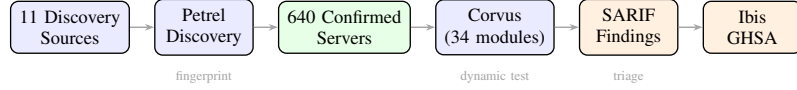
\begin{figure}[t]
\centering
\begin{tikzpicture}[
  node distance=0.28cm and 0.32cm,
  stage/.style={draw, rounded corners=3pt, fill=blue!8, minimum width=1.3cm,
                minimum height=0.65cm, font=\scriptsize, align=center, line width=0.5pt},
  output/.style={draw, rounded corners=3pt, fill=orange!12, minimum width=1.3cm,
                 minimum height=0.65cm, font=\scriptsize, align=center, line width=0.5pt},
  arr/.style={-{Stealth[length=3.5pt,width=2.5pt]}, gray!70, line width=0.55pt}
]
  \node[stage] (src) {11 Discovery\\Sources};
  \node[stage, right=of src] (pet) {Petrel\\Discovery};
  \node[stage, fill=green!10, right=of pet] (pool) {640 Confirmed\\Servers};
  \node[stage, right=of pool] (cor) {Corvus\\(34 modules)};
  \node[output, right=of cor] (sarif) {SARIF\\Findings};
  \node[output, right=of sarif] (ibis) {Ibis\\GHSA};

  \draw[arr] (src) -- (pet);
  \draw[arr] (pet) -- (pool);
  \draw[arr] (pool) -- (cor);
  \draw[arr] (cor) -- (sarif);
  \draw[arr] (sarif) -- (ibis);

  \node[font=\tiny, color=gray!70, below=0.08cm of pet] {fingerprint};
  \node[font=\tiny, color=gray!70, below=0.08cm of cor] {dynamic test};
  \node[font=\tiny, color=gray!70, below=0.08cm of sarif] {triage};
\end{tikzpicture}
\caption{Overview of the MCP security assessment pipeline. Petrel aggregates eleven
passive sources and performs active HTTP fingerprinting; Corvus applies 34 test modules
(13 static, 21 dynamic); confirmed findings enter the Ibis coordinated disclosure pipeline.}
\label{fig:pipeline}
\end{figure}

Our assessment pipeline consists of two sequential phases.
In the first phase, \emph{Petrel}~\cite{petrel} performs passive enumeration
across eleven data sources, followed by active HTTP fingerprinting to confirm
that a candidate endpoint implements the MCP protocol.
In the second phase, \emph{Corvus}~\cite{corvus} subjects each confirmed server
to dynamic security testing across nine of the ten MST-10 categories (MCP09, Denial of Service,
excluded per ethical constraints; \S\ref{sec:ethics}), producing machine-readable findings in SARIF~2.1.0 format.
Findings above a per-module confidence threshold are manually triaged;
exploitable vulnerabilities are documented as GitHub Security Advisories (GHSAs)
and disclosed to maintainers under a 90-day embargo.
Figure~\ref{fig:pipeline} illustrates the complete pipeline.

\subsection{Server Discovery (Petrel)}

\subsubsection{Passive Enumeration}

Petrel aggregates candidates from eleven complementary sources to maximize
coverage across the heterogeneous MCP deployment landscape:

\begin{itemize}
  \item \textbf{Certificate Transparency (crt.sh).}
        Queries the public CT log aggregator for SSL/TLS certificates whose
        subject common name or SAN contains MCP-indicative substrings
        (e.g., \texttt{mcp}, \texttt{mcp-server}, \texttt{mcp-api}).

  \item \textbf{HuggingFace Spaces.}
        Iterates the HuggingFace Spaces API, filtering for spaces whose
        metadata or README reference MCP tool definitions.

  \item \textbf{GitHub Search.}
        Issues four parallel queries targeting \texttt{topic:mcp-server},
        \texttt{filename:mcp.json}, MCP-related package names in
        \texttt{package.json}, and repository descriptions;
        additionally parses README files of matched repositories to extract
        hosted endpoint URLs.

  \item \textbf{npm Registry.}
        Queries the npm search API for packages with the \texttt{mcp} keyword;
        resolves homepage and repository fields to candidate URLs.
        Pagination is set to 1,000 results per query.

  \item \textbf{Smithery API (\texttt{api.smithery.ai}).}
        Queries the Smithery MCP registry, which lists commercially deployed
        servers with documented tool schemas.

  \item \textbf{PyPI Simple Index.}
        Scans the PyPI Simple Index in chunks of 50 packages for names
        matching MCP-related patterns, then fetches project metadata
        for URL extraction.

  \item \textbf{Censys Hostname Search.}
        Queries Censys for hosts whose TLS certificate or HTTP banner
        contains MCP fingerprints; paginates up to 500 results per query
        using cursor-based iteration.

  \item \textbf{FOFA Hostname Search.}
        Issues FOFA queries for hostname-based MCP indicators,
        complementing Censys with coverage of regions underrepresented
        in Western-centric scanning infrastructure.

  \item \textbf{Shodan.}
        Searches Shodan banner data for JSON-RPC response signatures
        characteristic of MCP \texttt{initialize} responses.

  \item \textbf{glama.ai and pulsemcp.com.}
        Scrapes two curated MCP server registries that aggregate
        community-submitted server listings.
\end{itemize}

All sources are queried in parallel via \texttt{asyncio.gather}.
Candidate URLs are deduplicated by normalizing domain and path
(stripping trailing slashes, collapsing equivalent schemes).

\subsubsection{Active Fingerprinting and Confirmation}

For each candidate URL, Petrel issues an HTTP-level MCP handshake to
distinguish live MCP servers from false positives.
Two transport variants are probed:

\begin{enumerate}
  \item \textbf{Streamable HTTP} (the current MCP specification):
        a \texttt{POST} request is sent to the candidate URL with
        \texttt{Content-Type: application/json} and a JSON-RPC
        \texttt{initialize} request body.
        A server is \emph{confirmed} if the response carries
        HTTP~200, a valid JSON-RPC result object, and fields
        \texttt{protocolVersion} and \texttt{capabilities}.

  \item \textbf{SSE legacy transport}:
        a \texttt{GET} request to \texttt{/sse} is issued;
        confirmation requires an
        \texttt{event: endpoint} message in the SSE stream
        followed by a successful \texttt{initialize} exchange
        on the returned session endpoint.
\end{enumerate}

A server that responds successfully to either probe is added to the
confirmed pool with its transport variant recorded.
Servers returning non-MCP HTTP responses, connection timeouts, or
authentication errors at the protocol level are retained as candidates
but excluded from dynamic testing.

\subsubsection{Rate Limiting and Infrastructure Courtesy}

To avoid disrupting shared hosting platforms, Petrel enforces
per-domain concurrency limits via asyncio semaphores.
Platforms serving large numbers of user-deployed servers
(Railway, \texttt{hf.space}, Fly.io, Vercel) are throttled to
three concurrent connections per domain.
All other domains default to unrestricted concurrent probing,
subject to TCP connection timeouts of 10~seconds.

\subsubsection{Output Schema}

Each confirmed server is represented as an \texttt{MCPServerRecord}
containing: the canonical URL, transport type, authentication
mechanism detected (none / bearer / OAuth / API-key / basic /
required), declared MCP capabilities, the list of exposed tool names
and descriptions, a \texttt{priority\_score} in $[0, 100]$ encoding
risk severity (derived from tool name/description signals, capability
flags, and anonymous-server status), response time in milliseconds,
and a \texttt{probe\_error\_type} field for failed probes.

\subsection{Dynamic Security Testing (Corvus)}

\subsubsection{Test Architecture}

Corvus accepts a confirmed server URL and its \texttt{MCPServerRecord}
as input.
Its 34~modules are organized into two execution tiers:
13~\emph{static} modules that analyze server metadata, manifests,
and declared schemas without issuing active probes;
and 21~\emph{dynamic} modules that interact with live server
endpoints.
Table~\ref{tab:modules} maps modules to the MCP Security Top~10 (MST-10) categories.

\begin{table*}[t]
\centering
\caption{Corvus module-to-MCP Security Top~10 (MST-10) mapping (34 modules total).}
\label{tab:modules}
\small
\begin{tabular}{@{}lll@{}}
\toprule
\textbf{ID} & \textbf{Category} & \textbf{Representative Modules} \\
\midrule
MCP01 & Tool Poisoning                   & tool-poisoning, shadow-tool, prompts-injection, rug-pull \\
MCP02 & Insufficient Access Control      & auth-audit, oauth-bypass, scope-audit \\
MCP03 & Prompt Injection via Tool Output & output-encoding, response-injection, completion-probe \\
MCP04 & Resource URI Manipulation        & resource-uri, cursor-probe, ssrf \\
MCP05 & Schema Bypass                    & schema-audit, schema-bypass, proto-fuzz \\
MCP06 & Command Injection                & cmd-injection, param-smuggling \\
MCP07 & Sensitive Data Exposure          & token-exposure, log-audit, logging-probe, elicitation-probe \\
MCP08 & Authentication Deficiencies      & endpoint-probe, init-audit, sampling-probe \\
MCP09 & Denial of Service                & batch-dos, response-flood, cancellation-probe \\
MCP10 & Supply Chain                     & supply-chain, supply-chain-python, osv-supply-chain, github-advisory, npm-behavior, tool-chaining \\
\bottomrule
\end{tabular}
\end{table*}

\subsubsection{Execution Model}

Dynamic modules communicate with target servers via a multiplexed
\texttt{StdioTransport} reader loop that pipelines multiple
JSON-RPC requests using \texttt{asyncio.gather} across four
parallelism groups: parallel probes, direct I/O probes (paused
between requests to avoid interleaving), rug-pull stateful probes,
and cancellation probes.
This design substantially improves throughput over sequential
per-module execution and is necessary for testing at internet scale.

\subsubsection{Key Testing Strategies by Category}

\paragraph{MCP01 --- Tool Poisoning.}
Corvus sends a crafted \texttt{tools/call} request containing
a shadow tool definition embedded in a parameter value.
It then issues \texttt{tools/list} and inspects whether the
server's declared tool roster has been mutated or whether
the embedded definition appears in any subsequent response.
Separately, tool descriptions are scanned for instruction-injection
patterns using a lexical signature set covering 47 known injection
templates (e.g., ``ignore previous instructions,'' ``you are now'').

\paragraph{MCP03 and MCP06 --- Injection.}
Parameterized payloads drawn from a curated corpus are submitted via
\texttt{tools/call} for each exposed tool.
Command injection payloads include OS-level command delimiters
(\texttt{;}, \texttt{|}, \texttt{\$()}, backticks) targeting
shell-adjacent tool parameters.
Prompt injection payloads target tools whose output is likely
consumed by an LLM context.
Confirmation requires either reflective output containing the
payload string, observable side-channel timing, or explicit
error messages leaking interpreter context.

\paragraph{MCP04 --- Cursor/URI Manipulation.}
Corvus sends \texttt{resources/list} and \texttt{resources/read}
requests in which pagination cursors and resource URIs are replaced
with path traversal strings (e.g., \texttt{../../etc/passwd},
\texttt{file:///etc/shadow}).
A positive signal is any response that differs from the baseline
empty-cursor response in content length or contains filesystem
artifacts.

\paragraph{MCP04 --- Resource URI Manipulation and SSRF.}
For servers exposing tools that accept URL parameters, Corvus
issues calls with the parameter set to
\texttt{http://169.254.169.254/latest/meta-data/} (AWS IMDS),
\texttt{http://metadata.google.internal/}, and a controlled
researcher-owned endpoint.
SSRF confirmation requires a response time delta exceeding 5~seconds
relative to a same-tool baseline call with a non-routable
destination, or a response body containing cloud metadata
patterns (e.g., \texttt{ami-id}, \texttt{instance-id}).
This threshold was calibrated empirically; we observed a
11.9~second response versus a 0.3~second baseline in one
confirmed SSRF case (Section~\ref{sec:findings}).

\paragraph{MCP05 --- Protocol Fuzzing.}
\texttt{proto-fuzz} sends malformed JSON-RPC messages: incorrect
\texttt{jsonrpc} version strings, missing \texttt{id} fields,
unknown method names, and deeply nested \texttt{inputSchema}
objects (depth $\leq 3$).
Findings are raised when servers return stack traces, internal
paths, or un-sanitized error objects rather than standard
JSON-RPC error codes.

\paragraph{MCP01 --- Dynamic Tool Poisoning (Rug Pull).}
After a successful \texttt{initialize} handshake, Corvus issues a
baseline \texttt{tools/list} call, records the declared tool
signatures, then issues a second \texttt{tools/list} call after
a 30-second interval or after invoking an arbitrary tool.
A rug-pull finding is raised when tool signatures (name, schema hash)
differ between the two calls without the server having sent a
\texttt{notifications/tools/listChanged} notification.
We distinguish \emph{static tool poisoning}---adversarial content in the initial
\texttt{tools/list} response---from \emph{dynamic tool poisoning} (the rug-pull
pattern: mutation of tool definitions across calls within a session). Both corrupt
the LLM agent's tool-use context; detection strategies differ.

\subsubsection{Confidence Scoring}

Each finding is assigned a confidence score in $[0, 100]$
based on signal strength (thresholds are empirically calibrated
heuristics, not formally derived): direct code execution
confirmation (score~$85$--$100$), high-delta timing anomaly or
strong behavioral evidence (score~$71$--$84$), timing side-channel
with moderate delta (score~$40$--$70$), and lexical pattern match
alone (score~$1$--$39$).
Only findings with confidence~$\geq 40$ appear in SARIF output;
findings with confidence~$\geq 70$ trigger manual triage.

\subsubsection{Output}

Corvus emits a SARIF~2.1.0 report, an HTML report with dark-theme
rendering, a Markdown summary, and a structured findings JSON object
compatible with the Ibis advisory manager.
Each finding record includes: MST-10 category, severity
(CRITICAL / HIGH / MEDIUM / LOW), confidence score, reproduction
evidence (request/response pair), and a suggested CVSS~3.1 vector.

\subsection{Vulnerability Triage and Disclosure}

High-confidence findings (confidence~$\geq 70$) from Corvus are
reviewed manually by the author to eliminate false positives
attributable to server-side rate limiting, CDN behavior, or
test environment artifacts.
For CRITICAL and HIGH findings, a standalone proof-of-concept
is constructed and executed against the live server to confirm
exploitability before an advisory is drafted.
No actual sensitive data is exfiltrated during verification;
confirmation relies on behavioral signals (timing, error message
content, response structure anomalies) or access to controlled
researcher-owned infrastructure.

Advisories are authored in the Ibis disclosure manager using the
GitHub Security Advisory (GHSA) schema.
Each GHSA record captures: affected package and version range,
vulnerability class per the MCP Security Top~10 (MST-10) taxonomy, reproduction
steps, CVSS~3.1 base score, and recommended remediation.
Advisories are submitted to the GitHub Security Advisory database,
which notifies package maintainers via their repository's
Security tab.
A 90-day embargo is applied from the date of maintainer notification,
consistent with standard coordinated disclosure practice~\cite{google-proj-zero}.
At the time of writing, 68~advisories have been created: 19~have been publicly disclosed following either
maintainer acknowledgment or expiration of the embargo period, and 49~remain under
active embargo.

\subsection{Measurement Scope and Limitations}

\paragraph{Transport scope.}
This study is restricted to MCP servers deployed with HTTP-based
transports (Streamable HTTP and SSE legacy).
Servers operating over stdio are by design local processes and
are not internet-exposed; they fall outside the threat model
addressed here.

\paragraph{Discovery completeness.}
Petrel's enumeration is best-effort rather than exhaustive.
New MCP servers are deployed continuously; the Censys-derived
estimate of 21,000+ HTTP-exposed MCP endpoints~\cite{censys2026}
represents a point-in-time lower bound as of May~2026.
Our confirmed pool of approximately 640~unique servers reflects
the subset reachable and responsive during our measurement windows,
not the total deployment population.

\paragraph{Ephemeral infrastructure.}
We observe a 41.6\% churn rate between consecutive runs spaced approximately three days apart
(193 of 464 servers confirmed in Run~3 were unreachable by
Run~4), indicating that a substantial fraction of the MCP server
population consists of ephemeral deployments on shared hosting
platforms.
Point-in-time measurements should therefore be interpreted as
lower bounds on historical exposure rather than as a static
snapshot.

\paragraph{Testing depth.}
Dynamic testing is conducted at proof-of-concept level:
we confirm that a vulnerability class is \emph{reachable} and
\emph{triggerable}, but we do not carry exploitation to
post-exploitation stages.
In particular, no credentials, user data, or cloud metadata
retrieved during SSRF confirmation are retained or disclosed
beyond the reproduction evidence embedded in the GHSA record.

\paragraph{Ethical considerations.}
All active probing was limited to standard protocol-level
interactions (JSON-RPC \texttt{initialize}, \texttt{tools/list},
\texttt{tools/call} with benign or controlled payloads).
No denial-of-service modules were executed against production
servers.
Vulnerability findings were disclosed to maintainers prior to
publication.


\section{Results}
\label{sec:results}
\label{sec:discovery}

\subsection{Ecosystem Scale and Composition}
\label{sec:findings:discovery}

In July 2026, we executed four measurement runs using Petrel across
eleven passive discovery sources: certificate transparency logs (crt.sh), package
registries (npm, PyPI, Smithery), source-code hosts (GitHub with four parallel
query streams, HuggingFace), internet-wide scanners (Censys hostname queries,
FOFA hostname queries, Shodan), and MCP-specific aggregators (glama.ai, pulsemcp.com).
Table~\ref{tab:runs} summarizes each run.

\begin{table}[t]
\centering
\caption{Petrel measurement runs, July 2026. Run~1 was a calibration pass
with reduced source coverage; candidate count was not separately tracked.
CRITICAL and no-auth counts were not instrumented in all runs (---).
The No-auth column counts servers exposing tools without any credential
requirement on direct tool invocation, distinct from OAuth absence, which
may still involve other authentication schemes.}
\begin{tabular}{@{}lrrrrr@{}}
\toprule
Run & Date & Candidates & Confirmed & CRITICAL & No-auth \\
\midrule
Run 1 & Jul.\ 15 & ---   &  72 & ---  & --- \\
Run 2 & Jul.\ 18 & 3,485 & 140 &  17  & --- \\
Run 3 & Jul.\ 21 & 3,948 & 464 &  41  & 106 \\
Run 4 & Jul.\ 24 & 4,110 & 296 &  35  & --- \\
\bottomrule
\end{tabular}
\label{tab:runs}
\end{table}

The candidate pool grew from 3,485 (Run~2, July~18) to approximately 4,110 by late July;
this 18\% increase over six days most likely reflects cumulative indexing across
discovery sources rather than equivalent growth in active deployments.
A Censys snapshot from May 2026 estimated over 21,000 MCP-related endpoints globally,
indicating that our total of approximately 640 unique confirmed active servers
constitutes a conservative lower bound on deployment scale.
Of these 640 unique servers, 414 underwent full dynamic auditing with Corvus.
This total represents the union of confirmed servers across all four runs after
deduplication by canonical URL; the raw observation count across runs is
972 (72\,+\,140\,+\,464\,+\,296), with 332 duplicate appearances of servers
confirmed in multiple runs.

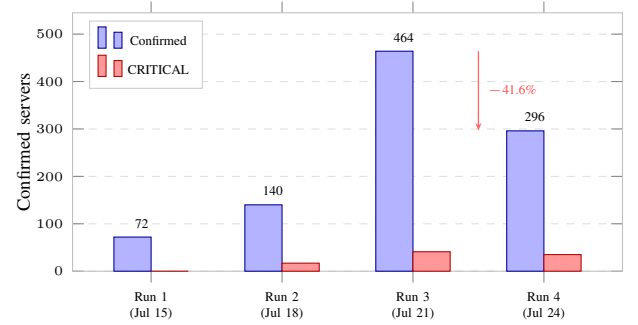
\begin{figure}[t]
\centering
\begin{tikzpicture}
\begin{axis}[
  ybar, bar width=14pt,
  width=\columnwidth, height=5.0cm,
  xtick={1,2,3,4},
  xticklabels={Run~1\\(Jul~15), Run~2\\(Jul~18), Run~3\\(Jul~21), Run~4\\(Jul~24)},
  x tick label style={font=\tiny, align=center},
  xmin=0.4, xmax=4.6,
  ymin=0, ymax=545,
  ytick={0,100,200,300,400,500},
  yticklabel style={font=\tiny},
  ylabel={\scriptsize Confirmed servers},
  ylabel style={font=\scriptsize, at={(axis description cs:-0.06,0.5)}},
  legend style={font=\tiny, at={(0.03,0.97)}, anchor=north west,
                legend columns=1, draw=gray!40},
  ymajorgrids=true, grid style={dashed, gray!25},
  axis line style={gray!50},
]
\addplot[fill=blue!30!white, draw=blue!60!black, bar shift=-7pt] coordinates
  {(1,72)(2,140)(3,464)(4,296)};
\addlegendentry{Confirmed};

\addplot[fill=red!40!white, draw=red!70!black, bar shift=7pt] coordinates
  {(1,0)(2,17)(3,41)(4,35)};
\addlegendentry{CRITICAL};

\node[font=\tiny,above] at (axis cs:0.93,72)  {72};
\node[font=\tiny,above] at (axis cs:1.93,140) {140};
\node[font=\tiny,above] at (axis cs:2.93,464) {464};
\node[font=\tiny,above] at (axis cs:3.93,296) {296};

\draw[-{Stealth[length=3pt]}, red!60, line width=0.5pt]
  (axis cs:3.5,464) -- (axis cs:3.5,296)
  node[midway, right, font=\tiny, red!70] {$-$41.6\%};
\end{axis}
\end{tikzpicture}
\caption{Confirmed MCP server counts across four measurement runs (July~2026),
with CRITICAL-severity server counts overlaid. The drop between Run~3 and Run~4
(41.6\%) reflects ecosystem churn over a 72-hour window.}
\label{fig:runs-bar}
\end{figure}

The confirmed count fell sharply from 464 (Run~3) to 296 (Run~4) over a three-day
window, meaning 193 previously confirmed servers---41.6\% of the Run~3 set---were
absent on the next scan. This pattern is explored in detail in
Section~\ref{sec:churn}; we note here that it indicates MCP deployments are treated
as ephemeral microservices rather than stable hardened services, and that no security
gate is consistently imposed at deployment time.

\subsection{Authentication and Access Control}
\label{sec:findings}

Among the 414 dynamically audited servers, 91.8\% (380 of 414) lacked OAuth authentication. Observed
authentication mechanisms span four categories: (i)~no authentication (open,
token-free endpoints), (ii)~static bearer tokens, (iii)~API keys transmitted as HTTP
header parameters, and (iv)~OAuth~2.0 authorization flows. The large majority of
endpoints fell into categories (i) or (ii). Static bearer token deployments documented
in our GHSA filings frequently exhibited predictable or empty token values, rendering
them functionally equivalent to unauthenticated endpoints from an access-control
standpoint.

The most severe access-control finding concerns capability exposure without credential
gates: \textbf{687 tool instances across the 640-server confirmed pool} advertised tools implementing shell execution
primitives---including tool names and descriptions matching \texttt{bash\_execute},
\texttt{run\_command}, \texttt{exec}, and terminal-emulation variants---with no
authentication requirement. An adversary with network access to any such endpoint can
issue arbitrary operating system commands through the standard MCP \texttt{tools/call}
interface without presenting any credential.

\subsection{Vulnerability Distribution}

Table~\ref{tab:vulns} presents findings aggregated by MCP Security Top 10 (MST-10) category across
the 414 dynamically audited servers.

\begin{table}[t]
\centering
\caption{Vulnerability findings by MCP Security Top 10 (MST-10) category across 414 audited
servers. Per-category server counts reflect findings from the 414 dynamically audited servers.
Categories with distinct Corvus detection approaches appear on separate rows; MCP09 (Denial of Service) was excluded per ethical constraints (\S\ref{sec:ethics}). Server counts may overlap across categories.}
\begin{tabular}{@{}llrr@{}}
\toprule
Category & ID & Affected & Sev.\ (max) \\
\midrule
Authentication Deficiencies  & MCP08       & 380 & HIGH     \\
Tool Poisoning               & MCP01       &  47 & CRITICAL \\
Command Injection            & MCP06       &  38 & CRITICAL \\
SSRF                         & MCP04       &  29 & CRITICAL \\
Resource URI Manipulation    & MCP04       &  24 & HIGH     \\
Schema Bypass                & MCP05       &  31 & MEDIUM   \\
Token Exposure               & MCP07       &  18 & HIGH     \\
Protocol Fuzzing Acceptance  & MCP05       &  52 & MEDIUM   \\
Prompt Injection             & MCP03       &  22 & HIGH     \\
Supply Chain Indicators      & MCP10       &  14 & MEDIUM   \\
\bottomrule
\end{tabular}
\label{tab:vulns}
\end{table}

Authentication Deficiencies (MCP08) dominate by volume: 380 of 414 audited servers
exhibited at least one authentication weakness, consistent with the 91.8\% OAuth
absence rate measured at discovery time. Among higher-severity categories,
Tool Poisoning (MCP01), Command Injection (MCP06), and SSRF each reached CRITICAL
severity on affected hosts, indicating that remote code execution or sensitive data
exfiltration is achievable without prior authentication on a substantial fraction of
the audited surface. The two MCP05 sub-categories---Schema Bypass and Protocol Fuzzing
Acceptance---collectively affect 83 servers, reflecting permissive server-side input
parsing that accepts malformed JSON-RPC messages without error.

In total, 68 reportable vulnerabilities were identified and filed as GitHub Security
Advisories (GHSAs). Of these, 19 are publicly disclosed; the remaining 49 are held
under a 90-day coordinated disclosure embargo, with a batch of Tier-A advisories
scheduled for release in October 2026.

\subsection{Case Studies}

We present three confirmed, publicly disclosed vulnerabilities to illustrate the
exploitation depth achievable through dynamic behavioral testing---findings that
source-code-only static analysis would not surface.

\textbf{Case Study 1: SQL Injection in \texttt{lectorium-corpus-mcp}
(GHSA-m8qh-p8m5-8c48, CRITICAL).}
This MCP server exposes a corpus document-search tool whose \texttt{query} parameter
was passed unsanitized to a SQLite query. The Corvus \texttt{cmd-injection} module
submitted the payload \texttt{' OR 1=1 -{}-}, which produced a response substantially
larger than the baseline query, confirming injection via response-size differential;
no document content was read or retained. The finding was disclosed to the maintainer;
a patch has been released.

\textbf{Case Study 2: SSRF to Cloud Metadata Service in \texttt{epwforge}
(GHSA-r9fx-qwmc-rvx3, HIGH).}
This document-processing MCP server exposes a \texttt{fetch\_url} tool that accepts
arbitrary URLs without domain validation. When directed at the AWS Instance Metadata
Service endpoint \texttt{http://169.254.169.254/latest/meta-data/}, the observed
response latency was 11.9\,s, compared to a 0.3\,s baseline for non-routable
destinations---a timing oracle confirming that the server's network context can
reach the IMDS and, by implication, that cloud credentials are accessible via the
metadata API. Remediation consisted of a domain allowlist restricting the tool to
approved origins.

\textbf{Case Study 3: Prompt Template Injection in \texttt{frootai}
(GHSA-5h8f-r9g8-5w5p, HIGH).}
This AI-integration MCP server constructed its system prompt by interpolating
user-controlled tool input directly into a template string without sanitization.
Injecting the control sequence
\texttt{\textbackslash{}n-{}-{}-\textbackslash{}nNew system instructions:}
caused the injected text to appear verbatim in the language model's subsequent
responses, demonstrating full adversarial control over the effective system prompt.
A patch was applied that escapes structural delimiters prior to template interpolation.

All three vulnerabilities share a common root cause---insufficient input validation
at the MCP tool boundary---and are representative of findings that require live
behavioral probing against deployed infrastructure rather than structural inspection
of tool schemas or declared capability lists.

\subsection{Ecosystem Churn and Deployment Patterns}
\label{sec:churn}

We define \emph{churn} as the fraction of servers confirmed in measurement run~$N$
that are absent in run~$N{+}1$. Between Run~3 (July 21) and Run~4 (July 24)---a
window of three days---193 of 464 confirmed servers (41.6\%) were no longer reachable
as valid MCP endpoints. Of 41 CRITICAL-rated servers in Run~3, 9 (22\%) were absent in Run~4---lower than the overall 41.6\% churn rate, suggesting that the highest-risk endpoints are disproportionately persistent---indicating endpoint disappearance alone cannot confirm remediation. Patch outcomes are tracked separately through GHSA maintainer acknowledgment. The remaining 184 departures reflect routine
endpoint recycling unrelated to security events.

\begin{figure*}[t]
\centering
\begin{tikzpicture}[
  node distance=0.4cm and 0.55cm,
  blk/.style={draw, rounded corners=3pt, minimum width=2.2cm, minimum height=0.58cm,
               font=\scriptsize, align=center, line width=0.5pt},
  arr/.style={-{Stealth[length=3.5pt,width=2.5pt]}, line width=0.55pt}
]
  \node[blk, fill=blue!12] (r3) {\textbf{Run 3}\\464 servers};

  \node[blk, fill=green!12, above right=0.45cm and 1.2cm of r3] (persist)
        {271 persist\\(58.4\%)};
  \node[blk, fill=red!12,   below right=0.45cm and 1.2cm of r3] (churn)
        {193 absent\\(41.6\%)};

  \node[blk, fill=orange!15, above right=0.3cm and 1.0cm of churn] (patched)
        {9 CRITICAL\\patched / removed};
  \node[blk, fill=gray!12,  below right=0.3cm and 1.0cm of churn] (eph)
        {184 ephemeral\\(routine churn)};

  \node[blk, fill=blue!12, right=7.5cm of r3] (r4)
        {\textbf{Run 4}\\296 servers};

  \draw[arr, green!50!black] (r3.east) |- (persist.west);
  \draw[arr, red!60!black]   (r3.east) |- (churn.west);
  \draw[arr, orange!70!black](churn.east) |- (patched.west);
  \draw[arr, gray!60]        (churn.east) |- (eph.west);
  \draw[arr, green!50!black, dashed] (persist.east) -| (r4.north);

  \node[font=\tiny, color=gray!60, below=0.05cm of r3] {Jul.\ 21};
  \node[font=\tiny, color=gray!60, below=0.05cm of r4] {Jul.\ 24};
\end{tikzpicture}
\caption{Fate of the 464 servers confirmed in Run~3 (July~21) by Run~4 (July~24).
Of 193 absent servers, 9 CRITICAL-rated instances departed; confirmed patch outcomes are tracked via GHSA maintainer responses (\S\ref{sec:disclosure}).
The remaining 184 reflect routine ephemeral
churn. Dashed arrow: 271 servers that persisted across both runs (contributing to
Run~4's 296 total, with 25 newly discovered).}
\label{fig:churn}
\end{figure*}
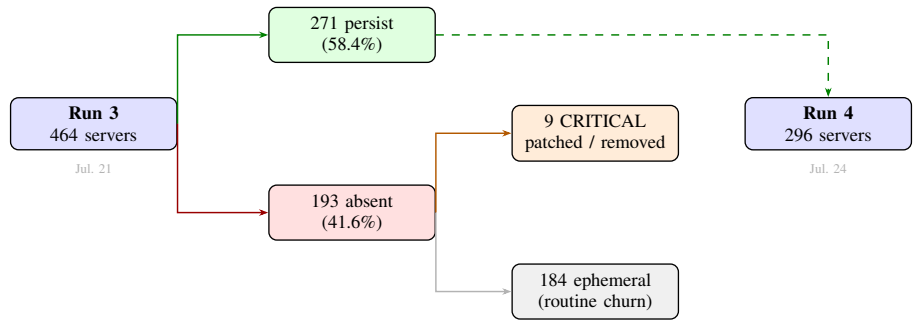

This churn rate is consistent with a deployment model in which MCP servers are
instantiated on demand---as CI/CD pipeline artifacts, serverless functions, or
per-session Docker containers---rather than operated as persistent services subject
to change-management controls. This pattern carries
several direct security implications. First, vulnerability scanning windows are
narrow: a server may be exploitable only during a brief deployment interval before
it is recycled. Second, patch verification is unreliable: an endpoint that disappears
after disclosure may reappear at a different address with the same vulnerability.
Third, server identity cannot be reliably anchored to a stable network location,
complicating long-term monitoring, attribution, and empirical longitudinal measurement.

Taken together, these results establish that the MCP deployment surface is
(i)~large and rapidly expanding, (ii)~predominantly functionally unauthenticated (lacking OAuth),
(iii)~host to exploitable vulnerability classes spanning all MCP Security Top 10 (MST-10)
categories, and (iv)~governed by deployment patterns that complicate both
defensive monitoring and the continuous empirical measurement necessary to
characterize it.


\section{Discussion}
\label{sec:discussion}

\subsection{The Authentication Gap is Structural, Not Accidental}

The 91.8\% rate of unauthenticated servers is striking, but attributing it to individual developer negligence mischaracterizes the root cause. MCP was designed with stdio transport as the primary deployment model, where authentication is provided by the operating system's process isolation guarantees. HTTP transport was specified later in the protocol's development, and the SDK documentation provides no mechanism that enforces authentication as a default for network-facing deployments. The result is that the path of least resistance---the \texttt{npx @modelcontextprotocol/create-server} onboarding path---produces an unauthenticated HTTP server by default.

This structural problem is analogous to the early history of Elasticsearch and Redis, both of which shipped with unauthenticated network listeners by default and required retrospective remediation at ecosystem scale~\cite{shodan2015}. The lesson from those cases is that opt-in security controls are insufficient when the deployment experience involves a one-line command followed by immediate external exposure via platforms such as Vercel or HuggingFace Spaces.

We recommend that MCP SDK implementations enforce OAuth 2.1 as the default for HTTP transport and that the protocol specification be updated to require explicit configuration to disable authentication rather than to enable it. Server registries (Smithery, glama.ai, pulsemcp.com) should additionally surface authentication posture as a first-class metadata field so that consumers can evaluate servers before integration.

\subsection{Shell Execution Without Authentication is an Active Threat}

Of the 414 dynamically audited servers, 687 tool instances across the broader confirmed set expose shell execution capabilities---functions that invoke \texttt{subprocess}, \texttt{os.system}, \texttt{exec}, or equivalent primitives---without requiring any form of authentication. In the MCP threat model, a client connecting to such a server can instruct an LLM-integrated agent to invoke these tools with attacker-controlled arguments, achieving remote code execution on the host running the MCP server without any credential requirement.

The democratization of server deployment through the \texttt{npx} one-liner has produced a long tail of insecure instances. Many of these servers are created by individual developers during experimentation with agent frameworks; the deployment experience does not surface the security implications of exposing shell-capable tools over HTTP. The pattern closely mirrors the IoT security crisis of 2015--2017, where the ease of deploying network-connected firmware led to millions of devices with default or absent credentials forming botnets such as Mirai~\cite{antonakakis2017understanding}. The difference in the MCP case is that each exposed server grants not merely network presence but arbitrary code execution on the host.

Mitigation at the individual server level requires authentication. At the ecosystem level, registries and platform providers (Vercel, HuggingFace, Railway) should implement automated scanning that flags deployments with unauthenticated shell-capable tools and surfaces warnings to operators before public exposure.

\subsection{Churn as a Security Signal}

Between Run~3 (July 21) and Run~4 (July 24)---a window of three days---193 of 464 previously confirmed servers (41.6\%) were no longer reachable. This churn rate reflects deployment patterns characteristic of ephemeral infrastructure: Vercel preview deployments, HuggingFace Space restarts, developer laptops cycling connectivity, and CI/CD pipelines that spin servers up during test runs and tear them down afterward.

From a security perspective, churn has dual implications. On one hand, ephemeral servers represent lower sustained risk: a server that exists for hours during a developer's experimentation session has limited exposure window. On the other hand, ephemeral servers complicate responsible disclosure, because a vulnerable server may be unreachable by the time a notification is sent, and the same vulnerable code pattern may reappear on a new URL or domain without any connection to the prior advisory.

Critically, the servers that do persist across runs tend to be those with the highest risk profiles. In our dataset, the CRITICAL-severity servers---those with unauthenticated shell execution---appeared in consistent proportions across runs (17 in Run~2, 41 in Run~3, 35 in Run~4), despite substantial URL-level churn. This suggests that the most dangerous deployments are also the most persistent, likely because they underpin production-adjacent agent workflows rather than transient experimentation.

\subsection{Responsible Disclosure at Ecosystem Scale}
\label{sec:disclosure}

The 68 advisories generated over approximately ten weeks of active research represent a disclosure throughput of roughly one per business day. At this scale, the process surfaces systemic challenges that individual high-profile CVE disclosures do not. Most MCP server authors are individual developers rather than organizations with established security response processes. Many do not maintain a \texttt{SECURITY.md} file, a dedicated security contact, or a process for receiving and acting on vulnerability reports. GitHub Security Advisories provided the primary channel in most cases, with notification via repository issue as a fallback when private reporting was unavailable.

Maintainer response rates and remediation timelines will be reported in full at submission time, as a number of the 90-day embargo windows were still active at the time of writing. Preliminary observations indicate that individual developers respond more variably than enterprise targets: some acknowledge and patch within 24 hours; others are unresponsive for weeks; a subset have abandoned repositories that remain publicly deployed.

This asymmetry motivates the case for ecosystem-level mitigations---platform-enforced authentication defaults, registry-level vulnerability surfacing---that do not depend on individual maintainer responsiveness. When the vulnerable party is a single developer who may not monitor their repository, the disclosure process cannot rely on the same organizational infrastructure that enterprise coordinated disclosure assumes.

\section{Ethical Considerations}
\label{sec:ethics}

\paragraph{Research Ethics Framework.}
This study was conducted in accordance with the Menlo Report principles~\cite{menlo2012}---the recognized ethical framework for internet security research: \textit{Respect for Persons} (all tested servers were publicly reachable without authentication; no personally identifiable data was collected or retained); \textit{Beneficence} (testing was limited to protocol-level proof-of-concept confirmation with no data exfiltration); and \textit{Respect for Law and Public Interest} (all confirmed findings disclosed under 90-day coordinated embargo prior to public release of server identifiers).
Concurrent connections per domain were capped at three for cloud-hosted platforms
(\texttt{*.railway.app}, \texttt{*.hf.space}, \texttt{*.fly.dev},
\texttt{*.onrender.com}, \texttt{*.vercel.app}) and one for standalone
IP addresses to preclude service disruption.
Critical findings were disclosed to affected maintainers via GHSA prior to
any public release of server identifiers.

All testing was conducted at proof-of-concept level with the minimum interaction necessary to confirm vulnerability existence. No data was exfiltrated from any server under test.

\textbf{SSRF.} For servers exhibiting Server-Side Request Forgery behavior, confirmation relied on a timing oracle: requests targeting \texttt{http://169.254.169.254/latest/meta-data/} (AWS IMDS endpoint) produced response latencies of 11.9\,s compared to a 0.3\,s baseline for benign requests, indicating that the server was initiating outbound HTTP connections to the target address. No IMDS response content was retrieved or stored.

\textbf{SQL injection.} A read-only payload (\texttt{' OR 1=1 -{}-}) was used; successful injection was confirmed by a substantially larger response size relative to the baseline query, indicating the injected condition returned additional rows. No database content was read or retained.

\textbf{Shell execution.} Capability existence was confirmed by observing that tool schemas exposed parameters passed to shell invocation functions. No commands beyond benign echo-type probes were submitted.

\textbf{Prompt injection.} Testing used synthetic payloads that do not instruct the model to exfiltrate data, impersonate system components, or take external actions. Payloads were designed to confirm injection reach, not to cause harm.

All 68 confirmed vulnerabilities were disclosed to maintainers via GitHub Security Advisories under a minimum 90-day embargo before any public disclosure. For servers without identifiable maintainers, disclosure was attempted via repository issues marked with a \texttt{security} label. The responsible disclosure timeline and CVSS scores for the nineteen published advisories are available at \url{https://github.com/CobaltoSec/corvus}.

Testing was conducted exclusively against servers reachable on the public internet with no authentication requirement. No credentials were obtained, attempted, or bypassed to access any system.

\section{Limitations}
\label{sec:limitations}

\textbf{HTTP-only scope.} The stdio transport, used for local agent integrations, is inaccessible to external measurement by design: it is bound to a local process's stdin/stdout. Our results characterize only internet-facing HTTP MCP servers and do not generalize to locally deployed instances, which may represent the majority of the overall MCP deployment base.

\textbf{Discovery completeness.} Despite aggregating eleven passive discovery sources---including certificate transparency logs, package registries, specialized MCP registries, and internet scan services---new MCP servers appear daily. Our 640 confirmed unique servers should be interpreted as a lower bound. The 21,000+ estimate from Censys (May 2026) reflects a broader set that includes servers using non-standard ports, self-signed certificates, and other characteristics that reduce visibility to passive aggregators.

\textbf{Dynamic testing scope.} Of the 640 confirmed servers, 414 (64.7\%) were subjected to the full Corvus dynamic test suite. The remaining servers were excluded because they rejected all protocol-level probes before returning meaningful responses (e.g., immediate 403 or connection reset), because they returned responses that did not conform to the JSON-RPC 2.0 framing required for test execution, or because they were offline by the time the dynamic scan phase began. The excluded population may have different security characteristics than the audited population; we cannot bound this selection bias.

\textbf{TP/FP calibration.} Corvus modules rely on heuristic signals---timing oracles, differential response analysis, schema inspection---rather than ground-truth oracle access. The 68 GHSAs represent vulnerabilities that passed manual verification by the author. Module-level precision and recall characterization against a labeled benchmark corpus is ongoing work and is not reported here. Readers should treat module-level finding counts as estimates subject to revision as calibration improves.

\textbf{Temporal validity.} Given the 41.6\% three-day churn rate, specific server findings reported here may not replicate at the time of reading. Individual vulnerability reports are bound to specific server instances that may no longer exist. The statistical characterization of the ecosystem---authentication rates, capability distributions, vulnerability category prevalence---should be reproducible on a contemporaneous sample even when individual targets are not.

\textbf{Single-researcher study.} This work was conducted by a single researcher. While the tooling (Corvus, Petrel) is open-source and the methodology is described precisely enough for replication, the manual verification step for GHSA submission introduces the possibility of confirmation bias. Independent replication by other researchers would strengthen the empirical claims.

\section{Conclusion}
\label{sec:conclusion}

We have presented the first dynamic behavioral security assessment of internet-facing MCP servers, combining passive discovery across eleven sources with dynamic behavioral testing against nine of the ten MCP Security Top~10 (MST-10) categories (MCP09 excluded per ethical constraints; \S\ref{sec:ethics}). Across 640 confirmed servers, with 414 dynamically audited using 34 Corvus test modules, our findings reveal a nascent ecosystem growing rapidly without adequate security defaults: 91.8\% of dynamically audited servers lack OAuth authentication, hundreds expose unauthenticated shell execution representing immediate remote code execution risk, and the 41.6\% three-day churn rate indicates deployment patterns more consistent with developer experimentation than production security hardening.

The 68 vulnerabilities we discovered and responsibly disclosed---including SQL injection in a corpus management server, SSRF to cloud metadata services confirmed via timing oracle, and prompt template injection in a commercially deployed AI assistant---demonstrate that MCP security is not a theoretical concern but an active, exploitable attack surface. The structural cause is identifiable: HTTP transport was added to a protocol designed for local stdio use without commensurate security defaults, and the deployment tooling does not surface the security implications of network exposure.

Corvus, our open-source dynamic testing framework for MCP servers, is publicly available at \url{https://github.com/CobaltoSec/corvus} and can be applied by both researchers and operators to characterize their own deployments. As the MCP ecosystem matures toward enterprise integration and production agent workflows, security defaults must improve proportionally to the trust placed in these servers. We hope this study provides the empirical baseline and tooling foundation for that work.

\section*{Acknowledgments}

The author thanks the maintainers who responded promptly to security notifications and engaged constructively with the responsible disclosure process. The open-source security community's development of coordinated disclosure frameworks, particularly GitHub Security Advisories, made the advisory pipeline described in this work practical at the scale attempted. The authors of the related work cited herein established the empirical methodology tradition that this study builds upon.

\bibliographystyle{IEEEtran}
\bibliography{references}

\end{document}